\documentstyle[pra,aps,floats]{revtex} 
\begin{document}
\newcommand{\beq}{\begin{eqnarray}} 
\newcommand{\eeq}{\end{eqnarray}}
\twocolumn[\hsize\textwidth\columnwidth\hsize\csname
@twocolumnfalse\endcsname

\title{
Lowest threshold visibility for testing local realistic theories 
}
\author{Gyula Bene}
\address{ 
Institute for Theoretical Physics, E\"otv\"os University,
     P\'azm\'any P\'eter s\'et\'any 1/A, H-1117 Budapest, Hungary
 }
\date{\today}
\maketitle

\begin{abstract}
Analytical and numerical arguments are presented in case of a pair
of two-state systems in a singlet state that the threshold visibility
for testing Bell's theorem on the entire range of measurement
settings is 33.3 \%. It is also shown that no lower
treshold exists.  
\end{abstract}

\vskip2pc]
\narrowtext
Bell's theorem states that local realistic theories cannot 
give account of all the quantum mechanical correlations. The best known 
proofs are based on the 
Clauser-Horne \cite{Clauser-Horne} or the Clauser-Horne-Shimony-Holt \cite{CHSH}      
inequality. These inequalities are consequences of local realism, 
thus experimental demonstrations of their violation\cite{exp1},\cite{exp2}  
imply  
the necessity of a revision of at least one fundamental physical concept. 
 
This far reaching consequence motivates the continued search for 
new methods for formulating Bell's theorem and testing local realism. 
A recent direction is to exploit the full angular dependence of the 
quantum mechanical correlations, which is equivalent with the case of  
infinitely many apparatus settings at each side of the experiment 
\cite{Zuk1}-\cite{Zuk2}. A standard Bell type experiment is assumed, 
i.e., a source emits two spin-half particles in a maximally entangled 
(singlet) state, so that 
the particles do not interact after the emission. Subsequently,  
the projection of the spin at a direction $\vec{a}$  
is measured on the first particle and  
the projection of the spin at a direction $\vec{b}$  
is measured on the second particle ($\vec{a}$, $\vec{b}$ are 
unit vectors).  
In a real experiment one has losses and thus  
the joint probability of having the results $m$ and $m'$ ($=\pm 1$) is 
\beq  
P_{QM}(m, m';\vec{a}, \vec{b})=\frac{1}{4}(1-m m' V \vec{a}\vec{b})\label{cqm} 
\eeq 
Here  
$0\le V \le 1$ stands for the visibility. It is an important question 
which visibility suffices to falsify the premises of local realism 
in case of infinitely many apparatus settings.   
In a recent paper\cite{Zuk2} Zukowski showed that the threshold  
visibility is not larger than $3/4$. This improved the previous estimates 
$8/\pi^2$ \cite{Zuk1} (derived for the case of coplanar settings)  
and $\pi/4$\cite{Gisin}. In the present paper we argue that the  
threshold visibility is actually $1/3$, and no lower value is possible. 
Throughout we adhere to the notations of Ref.\cite{Zuk2}. 
 
Local realism implies that the above joint probability is of the form  
\beq 
P_{HV}(m, m';\vec{a}, \vec{b})
=\int_\Lambda d\lambda \rho(\lambda) 
P_A(m|\vec{a}, \lambda) P_B(m'|\vec{b}, \lambda) \label{chv} 
\eeq 
 
where $\lambda$ stands for the 'hidden variable' characterizing the 
common past of the particles and $\Lambda$ for the space of its 
allowed values. The question is whether the actually observed  
joint probability (\ref{cqm}) admits the representation (\ref{chv}). 
We shall present an analytical and a numerical argument.  
The analytical treatment gives a precise value for the threshold 
visibility, but involves some additional assumptions concerning 
the nature of the hidden variable. In case of the numerical treatment 
we do not rely upon these additional assumptions, but the resulting 
value of the threshold visibility is determined only up to a finite 
accuracy. The analytical result coincides with the numerical one 
within this accuracy.     
 
In case of the analytical argument we make the following three 
assumptions: 
\begin{enumerate} 
\item  
As we have identical particles,
\beq 
P_A(m|\vec{n}, \lambda)=P_B(-m|\vec{n}, \lambda)\;.\label{a1} 
\eeq 
\item As $P_A(m|\vec{n}, \lambda)$ is a scalar, rotational 
invariance suggests that $\lambda$ is actually a unit vector, 
$\vec{\lambda}$ and  
\beq 
P_A(m|\vec{n}, \lambda)=f(m\vec{n} \vec{\lambda})\;.\label{a2} 
\eeq 
\item  
Rotational invariance also requires that there is no 
distinguished direction, thus 
\beq 
\rho(\lambda) d\lambda=\frac{1}{4\pi}d\Omega\;,\label{a3} 
\eeq  
where $d\Omega$ stands for an infinitesimal spatial angle around 
the direction $\vec{\lambda}$. 
\end{enumerate} 
 
Let us represent the function $f(x)$ (cf. Eq.(\ref{a2}), note that $-1\le x\le 1$) as 
a series of Legendre polynomials: 
\beq 
f(x)=\sum_{j=0}^\infty c_j P_j(x)\label{Lp}  
\eeq 
Here $c_j$-s are real coefficients. 
 
Inserting Eqs.(\ref{a1})-(\ref{Lp}) into Eq.(\ref{chv}) and using 
the identity 
\beq 
\int P_j(\vec{a} \vec{\lambda}) P_k(\vec{b} \vec{\lambda}) d\Omega 
=\delta_{jk}\frac{4\pi}{2 j+1}P_j(\vec{a} \vec{b})   
\eeq 
we have 
\beq 
P_{HV}(m, m';\vec{a}, \vec{b}) 
=\sum_{j=0}^\infty \frac{c_j^2}{2 j+1} P_j(-m m' \vec{a} \vec{b})  \label{chvLp} 
\eeq 
On the other hand, Eq.(\ref{cqm}) can be written as 
\beq 
P_{QM}(m, m';\vec{a}, \vec{b})=\frac{1}{4}P_0(-m m' \vec{a} \vec{b})
+\frac{V}{4} P_1(-m m' \vec{a} \vec{b})\label{cqmLp} 
\eeq  
Comparing this with Eq.(\ref{chvLp}) we get that $c_0$ and $c_1$ 
are the only nonzero coefficients and 
\beq 
c_0=\pm\frac{1}{2}\label{c0}\\ 
c_1=\pm\frac{\sqrt{3V}}{2}\;.\label{c1} 
\eeq 
If $\vec{a}$ and $\vec{\lambda}$ are perpendicular,  
$P_1(-m m' \vec{a} \vec{\lambda})$ 
is zero, thus the positivity of $P_A(m|\vec{a}, \lambda)$ requires 
(cf. Eqs.(\ref{a1}), (\ref{Lp})) that in Eq.(\ref{c0}) the positive 
sign must be chosen. If $\vec{\lambda} =\pm \vec{a}$, then  
$P_1(-m m' \vec{a} \vec{\lambda})=\pm 1$,  
thus the positivity of $P_A(m|\vec{a}, \lambda)$ requires 
that $c_0\pm c_1\ge 0$, i.e., by Eqs.(\ref{c0}), (\ref{c1}) 
\beq 
V\le \frac{1}{3}\label{tresh} 
\eeq 
As $|P_1(x)|\le 1$, Eq.(\ref{tresh}) 
ensures the positivity of $P_A(m|\vec{a}, \lambda)$,  
$P_B(m'|\vec{b}, \lambda)$ 
for any directions 
$\vec{a}$, $\vec{b}$, $\vec{\lambda}$.   
This readily implies that the threshold visibility is $1/3$, 
as below this value the observed joint probability (\ref{cqm}) 
admits a local realistic representation (\ref{chv}), while 
above this value the positivity of the conditional 
probabilities $P_A(m|\vec{a}, \lambda)$, $P_B(m'|\vec{b}, \lambda)$   
cannot be achieved. 
 
Although the assumptions (\ref{a1})-(\ref{a3}) look reasonable, 
one might suspect that the above result hinges upon these assumptions 
and perhaps without them  
the joint probability (\ref{cqm}) admits a local realistic representation  
even at higher visibilities. We present a numerical argument  
that it is not the case.  
 
The numerical procedure begins with choosing $N$ directions for $\vec{a}$
and $N$ directions for $\vec{b}$ and then equating Eqs.(\ref{cqm}) 
and (\ref{chv}). In the latter the integral is estimated by a sum,
i.e., the equation
\beq
\frac{1}{4}(1-m m' V \vec{a}_j\vec{b}_k)=\sum_{n=1}^M\rho_n
P_A(m|\vec{a}_j, n) P_B(m'|\vec{b}_k, n)
\eeq 
is to be solved. This is equivalent with
\beq
V \vec{a}_j\vec{b}_k=\sum_{n=1}^M\rho_n A_{j, n} B_{k, n}\label{num1}
\eeq   
where
\beq
A_{j, n}=1-2\;P_A(1|\vec{a}_j, n)\nonumber\\ 
B_{k, n}=2\;P_B(1|\vec{b}_k, n)-1
\eeq
which satisfy
\beq
|A_{j, n}|\le 1\nonumber\\ 
|B_{k, n}|\le 1\label{num2}
\eeq
and
\beq
\sum_{n=1}^M\rho_n A_{j, n}=0\nonumber\\ 
\sum_{n=1}^M\rho_n B_{k, n}=0\;.\label{num3}
\eeq
At the solution of Eq.(\ref{num1}) one can utilize the singular value
decomposition
\beq
\vec{a}_j\vec{b}_k=\sum_{i=1}^3 p_i U_{j, i} V_{k, i}\label{singval}
\eeq
where $U_{j, i}$ and $V_{k, i}$ are orthogonal matrices. Note that
there are only three nonzero singular values $p_n$, owing to the
three dimensionality of the vectors $\vec{a}_j$, $\vec{b}_k$.
A solution of Eqs.(\ref{num1})-(\ref{num3}) can be then obtained
in the following way:
\begin{enumerate}
\item One chooses three M dimensional vectors $\vec{q}_i$ 
and another three M dimensional vectors $\vec{t}_i$ and makes them
orthogonal to each other,
\beq
\vec{q}_i \vec{t}_j=\delta_{i,j}\label{ortog1}
\eeq
and to the vector $(\sqrt{\rho_1},\sqrt{\rho_2},... \sqrt{\rho_n})^T$,
i.e.
\beq
\sum_n \sqrt{\rho_n} (\vec{q}_i)_n=0\nonumber\\
\sum_n \sqrt{\rho_n} (\vec{t}_i)_n=0\;.\label{ortog2}
\eeq
\item Calculate
\beq
A_{j, n}'=\sum_{i=1}^3 U_{j, i} \sqrt{p_i}(\vec{q}_i)_n/\sqrt{\rho_n}\nonumber \\ 
B_{k, n}'=\sum_{i=1}^3 V_{k, i}\sqrt{p_i}(\vec{t}_i)_n/\sqrt{\rho_n}\label{main}
\eeq
\item A solution of Eq.(\ref{num1}) is given by
\beq
A_{j, n}=\sqrt{V}A_{j, n}' \nonumber\\ 
B_{k, n}=\sqrt{V}B_{k, n}' \label{ab}
\eeq
\item Eq.(\ref{num2}) implies that
\beq
1/\sqrt{V}=\max_{j,n}\left\{|A_{j, n}'|, |B_{j, n}'|\right\}\label{v}
\eeq
\end{enumerate}
Applying this scheme, a Monte-Carlo simulation has been used
to find the threshold visibility. The steps of the simulation are
the following:
\begin{enumerate}
\item Choose the unit vectors $\vec{a}_j$, $\vec{b}_k$ at random
\item Choose the vectors $\vec{q}_i$, $\vec{t}_i$ at random
and normalize them according to Eqs.(\ref{ortog1}), (\ref{ortog2}). 
\item Choose $\rho_n$ at random and normalize by
\beq
\sum_n \rho_n=1
\eeq 
\item Calculate $V$ from Eqs.(\ref{main}), (\ref{v}).
\item Change a component of $\vec{q}_i$, $\vec{t}_i$ or $\rho_n$ 
at random and repeat the calculation of $V$. If the resulting
value is larger than the previous one, keep the changes, otherwise
discard them.
\item After having found the maximal $V$ for fixed $\vec{a}_j$, $\vec{b}_k$,
change these vectors at random and find (by repeating the previous steps) the corresponding $V$ again. The minimum of these $V$ values is selected, because
at the corresponding setting for $\vec{a}_j$, $\vec{b}_k$ a local realistic representation of the observed joint probabilities at higher visibility cannot be given.   
\end{enumerate} 
Obviously, one is interested in the limit $N\rightarrow \infty$. For a finite
$N$ one has less restrictions, thus the resulting estimate for the
threshold visibility is higher. By increasing $N$ the estimates decrease monotonously. As for the number $M$, numerically it turned out that increasing its value makes the convergence slower but does not influence
the results for $V$. Hence one may set it to the minimum $M=4$.
The values for $N$ have been changed from 3 to $10^4$. The convergence
proved to be rather slow, e.g., the estimate for the
threshold visibility at $N=1000$ is still $0.37\pm 0.001$. The
final numerical result (extrapolation for $N\rightarrow \infty$) 
is $0.33\pm 0.03$. This is consistent with the previous analytical
result. 

It is instructive to compare these results with those one may obtain
from Bell's inequality and the Clauser-Horne-Shimony-Holt inequality.
Since these inequalities do not exploit the full angular dependence,
one gets a higher estimate for the threshold visibility. Note, however, that 
at the derivation of Bell's inequality
\beq
P_{HV}(1, 1;\vec{a}, \vec{b})+P_{HV}(1, 1;\vec{b}, \vec{c}) \ge P_{HV}(1, 1;\vec{a}, \vec{c})\label{bell}
\eeq
strict anticorrelation (i.e., $P_{HV}(1, 1;\vec{b}, \vec{b})=0$) implied by the singlet state is also assumed. Inserting Eq.(\ref{cqm}) into Eq.(\ref{bell})
we have
\beq
\frac{V}{2}\left
(3-(\vec{a}+\vec{c}-\vec{b})^2\right)\le 1\;.\label{bell2}
\eeq
The l.h.s. will be the largest when $\vec{a}+\vec{c}-\vec{b}=0$. This 
implies that the threshold visibility $V$ is not larger than $2/3\approx 0.667$,
as below this value Eq.(\ref{bell2}) is always satisfied, while
for a larger value of $V$ it may be violated for suitably chosen
directions $\vec{a}, \vec{b}, \vec{c}$. This value is smaller than 
the previously known smallest value $3/4=0.75$ \cite{Zuk2}, but here the additional
information about the properties of the singlet state has also been
utilized. This is equivalent with the assumption that in the special case of $\vec{a}=\vec{b}$ (cf. Eq.(\ref{cqm})) the visibility is 100\%. 
Using only the requirements of local realism together
with the full angular dependence we obtained above the much lower
threshold visibility $1/3\approx 0.333$. 

Let us consider now the Clauser-Horne-Shimony-Holt inequality
\beq
P_{HV}(1, 1;\vec{a}, \vec{b})-P_{HV}(1, 1;\vec{a}, \vec{b}') 
+P_{HV}(1, 1;\vec{a}', \vec{b})\nonumber\\
+P_{HV}(1, 1;\vec{a}', \vec{b}') 
-P_{HV}(1;\vec{a}')-P_{HV}(1;\vec{b})\le 0\label{CHSH}
\eeq 
Here $P_{HV}(1;\vec{a}')=\sum_{m'} P_{HV}(1, m';\vec{a}', \vec{b})$
and $P_{HV}(1;\vec{b})=\sum_{m} P_{HV}(m, 1;\vec{a}', \vec{b})$. 
Inserting Eq.(\ref{cqm}) into Eq.(\ref{CHSH}), the resulting
inequality can be cast to the form 
\beq
\frac{V}{2}\left((\vec{a}+\vec{b}'-\vec{b})^2+(\vec{a}'-\vec{b}'-\vec{b})^2-6\right)\le 2\;.\label{CHSH2}
\eeq
For fixed $V$, $\vec{b}$ and $\vec{b}'$ the l.h.s of inequality (\ref{CHSH2}) is the largest if $\vec{a}$ is parallel to $\vec{b}'-\vec{b}$ and
$\vec{a}'$ is antiparallel to $\vec{b}'+\vec{b}$. Then the lengths of
$\vec{a}$ and $\vec{b}'-\vec{b}$ add up to the length of 
$\vec{a}+\vec{b}'-\vec{b}$ (similarly in case of $\vec{a}'$). Thus,
denoting the angle between $\vec{b}$ and $\vec{b}'$ by $\varphi$,
we get from Eq.(\ref{CHSH2}) the inequality
\beq
2\sqrt{2}V\sin\left(\frac{\varphi}{2}+\frac{\pi}{4}\right)\le 2
\eeq
This readily implies that the l.h.s. is maximal (for a fixed $V$) 
if $\varphi=\frac{\pi}{2}$ and the threshold visibility is not larger
than $1/\sqrt{2}\approx 0.707$. Note that even this value is smaller than the previously obtained smallest value  $3/4=0.75$ \cite{Zuk2}. The CHSH inequality
relies only upon the assumption of local realism, but does not
exploit the full angular dependence. As we have seen, taking into account
the latter, too, a much lower value for the treshold visibility
can be deduced.

{\bf Acknowledgements}

This work has been partially supported by the Hungarian Aca\-demy of
 Sciences
 under Grant No. OTKA T 029752, T 031 724  and the J\'anos Bolyai Research Fello
wship.

\end{document}